\documentstyle[preprint,aps,epsfig]{revtex}

\tightenlines 
\begin{document}
\draft


\title{%
Structural and electronic properties of the sodium tetrasilicate glass 
Na$_{\mathbf 2}$Si$_{\mathbf 4}$O$_{\mathbf 9}$  from classical 
and {\it ab initio} molecular-dynamics simulations.
}

\author{%
Simona Ispas\footnote{Author to whom correspondence should be addressed:
ispas@ldv.univ-montp2.fr}, Magali Benoit, Philippe Jund and R\'emi Jullien
}

\address{
 Laboratoire des Verres, Universit\'e Montpellier II, 
         Place E. Bataillon, 34095 Montpellier, France 
}

\date{\today}
\maketitle

\begin{abstract}
The structure and the electronic properties  of a sodium tetrasilicate  
(Na${}_2$Si${}_4$O${}_9$) glass
were studied by combined Car-Parrinello and 
classical molecular-dynamics simulations. The glass sample was 
prepared using a method recently employed  in a study of a silica glass  
 [M. Benoit {\it et al}, Euro. Phys. J. B {\bf 13}, 631 (2000)]. 
First we  generated a NS4 glass  by classical molecular-dynamics 
and then we took it as the initial configuration 
of a first-principles molecular-dynamics simulation. 
In the {\it ab initio} molecular-dynamics simulation, the electronic structure 
was computed in the framework of the Kohn-Sham density functional theory within the 
generalized gradient approximation using a B-LYP functional. The Car-Parrinello dynamics
is remarkably stable during the considered trajectory and, as soon as it is switched on, 
some significant structural changes occur. The {\it ab initio}
description improves the comparison of the structural characteristics 
with experimental data, in particular concerning the Si-O and Na-O bond lengths. 
From an electronic point of view, 
we find that the introduction of the sodium oxide in the silica network lowers 
the band gap and leads to a highly non-localized effect on the charges of 
the network atoms. 

\end{abstract}


\newpage

\section{Introduction}
\label{sec:intro}

Over the past three decades, a large number of experimental and theoretical investigations 
has been carried out in order  to extract useful 
information about the structure and the physical properties of 
silicate glasses. These materials have stimulated great interest because the advances in 
understanding them are essential for progress in many 
scientific fields, such as glass and ceramics chemistry, 
electronics, earth sciences, or the confinement of industrial and nuclear wastes.
However, in spite of their evident importance, there are still important 
ambiguities in their characterization at a microscopic level.
This situation is mainly due to the  inherent structural disorder present in 
glasses which complicates  the understanding of the location of 
the different elements by 
comparison with crystalline materials.  

 In view of the current experimental findings and theoretical models, it is generally accepted
 that the structure of the most common, widely used and intensely studied glass,  pure amorphous silica,
is composed of SiO${}_4$ tetrahedra which are linked together in a 
continuous random network by strong bonds of bridging oxygens (BO), 
i.e. oxygen atoms bonded to two silicon atoms \cite{devine_88}. 
When a given amount of alkali (Li${}_2$O, Na${}_2$O or K${}_2$O), or alkaline 
earth (CaO, MgO) metal oxide  is incorporated, the  connectivity of this
 network is 
decreased by the breaking of BO bonds and the formation of non-bridging 
oxygen atoms (NBO) which are  bonded to only one silicon atom and weakly bonded 
to a cation modifier. Consequently the local environment around the silicon atoms changes 
depending on the number of bridging oxygen atoms lying within the first shell. 
This is commonly described by means of the notation $Q_n$ 
where $n$ is the number of BO atoms linked to a given Si atom (n=4,3,2,1,0). 
From a technological point of view, the introduction of an alkali metal oxide in silica
 is accompanied by a large reduction of the viscosity, an increase in the thermal expansion 
coefficient, and  a change in  the refractive index: these physical properties all depend on 
the bridging to non-bridging oxygen ratio in the glass \cite{mysen_88}.

In particular an enormous amount of work has been done on  binary sodium silicate glasses since 
almost all  commercial silicate glasses  and  geologic magma are based on this
family of compositions. They may also  serve as 
useful prototypes to interpret the structural and transport properties of more complicated  
silicate glasses and melts (see  for example 
\cite{brawer_77,greaves_81,greaves_85-00,brown_95,brawer_74,furukawa_81,mazzara_00} 
and references therein). 

In addition to physical experiments and to advances in techniques such as EXAFS, NMR or XPS,  
classical molecular-dynamics (MD) simulations have began to provide  some insight into 
the three-dimensional structure of these glasses. However one has to notice that, even 
though  reasonable results have been reported  
\cite{soules_81,tesar_87,huang_90,huang_91,melman_91,smith_95,oviedo_98,siona_kob}, 
this kind of   molecular dynamics studies utilizes an interatomic potential  generally 
fixed once and forall throughout the simulation.
 Moreover  it is difficult to adjust a classical potential functional form and parameters   
which are able to incorporate  both a covalent
 and an ionic character for the bonds existing in the real systems in order   
to obtain a genuine realistic description. However there exists an alternative approach, 
the {\it ab initio} molecular dynamics simulations, where  these drawbacks are, in principle, 
eliminated  as  the interatomic potential is updated at every simulation step by first principles
 calculations. The charge transfers between the NBO and 
the modifier ions can thus be correctly accounted for. 
Unfortunately one has to pay a price for these obvious advantages, namely   
a small system size and an  extremely limited maximum length of the trajectory.

In this article, we present the microscopic characteristics of a sodium tetrasilicate glass 
Na${}_2$Si${}_4$O${}_9$ (called hereafter NS4) derived from  an {\it ab initio} 
molecular dynamics simulation. Numerous experimental studies 
(NMR  \cite{dupree_84,emerson_89,maekawa_91,kummerlen_92}, 
neutron diffraction \cite{zotov_96,zotov_98_1,zotov_98_2}, 
Raman spectroscopies \cite{zotov_98_2,wolf_90}, XAFS \cite{houde-walter_93,greaves_95}, ...) 
have been carried out in order to elucidate the structure and the properties of this glass. 
The NS4 system has been also analyzed in  classical MD calculations 
\cite{tesar_87,huang_90,huang_91,melman_91,zotov_99} or 
in reverse Monte Carlo (RMC) fits \cite{zotov_98_2}. 
The attention given to this system may be justified by the fact that it can be used 
as a simple model for more complicated aluminosilicate and hydrous silicate glasses.

The present NS4 glass sample was generated using an approach originally employed  
for obtaining and studying a silica glass sample \cite{sio2_cp_bks} in two successive steps. 
First the liquid equilibration, the quench and the glass relaxation are performed 
classically using a model potential having the same functional form  as the so-called BKS potential
 (van Best {\it et al.} \cite{bks_sio2}) for pure silica.  In a second step, the resulting glass 
configuration is relaxed using the Car-Parrinello (CP) method \cite{car-parrinello_85,marx-hutter_00}.
 As in the silica study, a significant amount of CPU time is saved by adopting 
this strategy since the liquid equilibration, the quench and part of the low
temperature relaxation are carried out within the framework of classical MD.

The outline of the paper is as follows:  In Sec.\ref{sec:methods}, the methodology is briefly 
described and the details of the particular simulations performed here are
given.  In Sec.\ref{sec:results}, the main results are
presented, including structural properties of the
NS4 sample in Sec.\ref{sec:struct} as well as electronic properties
 in Sec.\ref{sec:electronic}. Section \ref{sec:concl}  gives the conclusions.


\section{Simulation details}
\label{sec:methods}

We started by generating a liquid sample at a temperature around 3500 K  
containing 90 atoms (24 silicon atoms, 54 oxygen atoms and 12 sodium atoms)
 confined in a cubic box of edge length 10.81 \AA \ corresponding 
to the experimental NS4 mass density of 2.38 g/cm${}^3$ \cite{bansal_doremus}. 
The initial configuration of the NS4 liquid was obtained by melting at 3500 K a  
$\beta$-cristobalite  crystal in which    SiO${}^{-2}_4$ tetrahedra were
 replaced  randomly  by Na${}_2$O${}^{-2}_3$  groups. 

For this system we performed classical molecular 
dynamics simulations  within  
the microcanonical ensemble, with periodic boundary conditions. 
 As mentioned above, we have chosen  the interaction potential to be
 a generalization of the BKS potential \cite{bks_sio2} successfully applied for 
silica \cite{sio2_bks} and which was initially developed by Kramer {\it et al.} 
\cite{bks_na_kramer} in order to study  zeolites. Recently Horbach  
{\it et al.} \cite{siona_kob} have  adjusted  it  for sodium silicate studies  by 
introducing an additional short range term for the sodium ion in order to reproduce 
the experimental mass density and the structure factor of the sodium disilicate glass. 

\begin{figure}
\epsfxsize=280pt{\epsffile{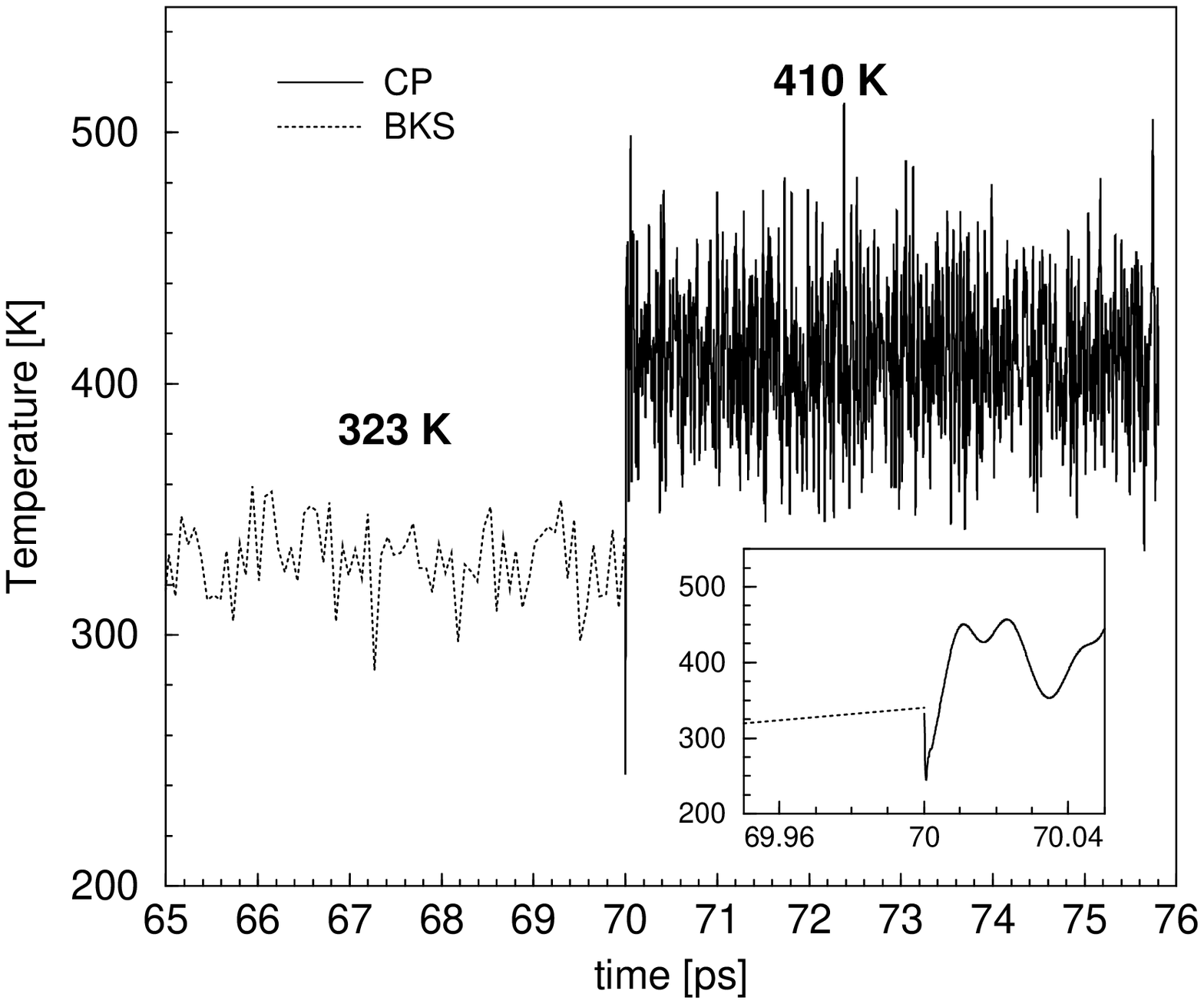}}
\caption{\label{evol-temp} Ionic temperature evolution during the last 5 ps of 
the classical MD simulation (dotted line) and the full Car-Parrinello MD simulation (solid line). 
In the inset,  a zoom of the beginning of the Car-Parrinello MD simulation is depicted.}
\label{fig:evol-temp}
\end{figure}

The NS4 liquid was equilibrated during 70 ps and then cooled from 3500 K to $\approx $ 300 K 
at a quench rate of 5 $\times$ 10${}^{13}$ K/s, using a time step of 0.7 fs. 
The cooling procedure used here is identical
to the one used in former studies of SiO2$_2$  \cite{sio2_cp_bks,sio2_bks} 
and it has been shown that it gives access to structural, dynamical and thermal properties in 
good agreement with experiments.
The resulting glass was relaxed  during another 70 ps and then the  final  classical 
configuration was used as starting point for the Car-Parrinello  {\it ab initio} simulation. 
Immediately after switching on the CP dynamics we observed a temperature
 jump  similar to  the one reported in the silica study  \cite{sio2_cp_bks}, 
which was done  following the same approach (see Fig. \ref{evol-temp} where the evolution 
of the ionic temperature is depicted  for the end of the classical MD trajectory 
and the full subsequent CP trajectory). Hence in the CP simulation after a very short relaxation 
(see the inset of the Fig. \ref{evol-temp}), the system heats up and then 
its temperature remains stable around an average value of $\approx$ 410 K 
(87 K higher than that of the classical MD simulation).

The CP simulation was performed with the {\it ab initio} MD code, CPMD developed in 
Stuttgart \cite{cpmd_code}. In the {ab initio} MD simulations, the 
electronic structure was treated via the Kohn-Sham (KS) formulation 
of density functional theory \cite{kohn_sham} within the generalized gradient 
approximation employing the B-LYP functional \cite{becke88,lyp}. 
The KS orbitals were expanded in a plane wave basis at the $\Gamma$-point of the 
supercell up to an energy cutoff of 70 Ry. Core electrons were not treated 
explicitly but replaced by atomic pseudopotentials of the Goedecker type
for silicon \cite{SG} and the Troullier-Martins type for oxygen \cite{MT},
 while for sodium a Goedecker type semi-core pseudopotential was employed. 
Preliminary tests have been performed to check the adequacy  of the choices of the 
pseudopotentials, exchange and correlation functionals and energy cutoff value. 
The choices of pseudopotentials for Si and O and the energy cutoff value led to
a good convergence for the structural properties at 0 K in cristobalite 
and $\alpha$-quartz. The  sodium pseudopotential and exchange and correlation functionals 
are justified by bond length estimations carried out on Na${}_2$ and Na${}_2$O molecules 
and which have been found to be close to  
experimental values \cite{huber_herzberg,na_liquid}  or more sophisticated
 theoretical  calculations available in the literature \cite{elliott_ahlrichs}.

The CP relaxation was performed for $\approx$ 6 ps, with a time step of 5.5 a.u.
(0.13 fs) and a fictitious electronic mass of 800 a.u. 
used for the  integration of the equations of motion. 
The data have been averaged over the last 5 ps of the trajectory.

The procedure described in the above paragraphs has been carried 
out a second time following exactly the same steps and using the same 
parameters for both BKS and CP simulations, but starting from a different BKS liquid sample.
 Thus we have generated a second NS4 glass sample which was once again 
remarkably stable during the CP trajectory after   an identical
temperature jump showed up at the beginning of the CP simulation. 
Indeed this second sample heats up from an average 
temperature of $\approx $ 315 K at the end of the BKS relaxation to 
an average CP temperature of $\approx$ 395 K. Noting that the temperature
jumps were of the same order for both samples, we may explain this phenomenon 
as being due to the fact that under CP, the samples relaxed towards a potential minimum lower that
the one calculated during the BKS trajectories
and this supplementary potential  energy has been transformed into ionic kinetic energy. 
In the next sections, we will discuss the structural properties obtained by averaging 
the data collected for  both samples during their BKS and  CP simulations.


\section{Results and discussion}
\label{sec:results}


\subsection{Structure}
\label{sec:struct}

In this subsection, we present   the structural analysis of both NS4 samples 
 by discussing  commonly interrelated features  like the pair
correlation functions, bond angle distributions, coordination numbers, 
Q-species distributions, and bridging to non-bridging oxygen ratio. 
These topics have been investigated for both classical 
and CP samples in order to understand 
 how much the local arrangement of network formers and 
modifiers is changed when 
the  system is treated correctly from a quantum mechanical point of view.

We start by remarking  that both CP and BKS simulations give the same NBO concentration 
percentage (22.22\%) equal to  the sodium-ion concentration which is the  value 
 predicted theoretically  and confirmed by experiments \cite{maekawa_91,jen_89}.
We have identified the NBO atoms  as being either the oxygen atoms having one Si and one Na atom 
as nearest neighbours or the  oxygen atoms having only one  Si at a distance below 
 a Si-O bond cutoff (taken to be 2.0 \AA, i.e.  where  the Si-O pair correlation function 
becomes zero after the first peak - see below) and we have obtained the same results. 
Analyzing the neighbourhood of the NBO atoms, we have systematically found 
that each of these atoms has two Na atoms at almost equal distances,  as  second and third 
nearest neighbours. We have also considered the nearest-neighbours of each Si atom 
and we have found that all Si atoms were coordinated by 4 oxygens  during both runs.  

Further, once the NBO atoms have been identified, we have determined the Q-species distributions 
which are equal to  54.2$\%$ Q${}_4$, 41.6$\%$ Q${}_3$, 4.2\% Q${}_2$ for the first sample and 
to 58.3$\%$ Q${}_4$, 33.3$\%$ Q${}_3$, 8.3\% Q${}_2$ for the second one.  
We note that the switching on of the CP description did not change these percentages.
For the NS4 glass,  previous ${}^{29}$Si magic-angle spinning nuclear
magnetic resonance experiments reported the following results~: 55$\%$ Q${}_4$, 
40$\%$ Q${}_3$, 5\% Q${}_2$ by Emerson et al. \cite{emerson_89} and 50$\%$ Q${}_4$, 
48$\%$ Q${}_3$, 2\% Q${}_2$  by Maekawa et al. \cite{maekawa_91}. Our simulation results
for the first sample compare relatively favorably with the experimental data, while, 
for the  second one, there is an excess of Q${}_4$ and Q${}_2$ species.  Nevertheless, 
in spite of the limited size of the simulated glass and of the ultrafast quench rate, 
we may point out that there is good overall agreement from the experimental 
and the simulation data for the Q-species distribution.

From a structural point of view, we note that some significant effects appear 
as soon as the CP dynamics starts and all of them take place on a very short period of time
($\approx$ 0.1 - 0.2  ps) which can be related to the stabilization time of the 
temperature. We will examine, in the next paragraphs, each of these effects occurring 
when switching from the classical to the {\it ab initio} description 
and we will discuss the corresponding average structural properties.

\subsubsection{Mean angles and angular distributions}

\begin{figure}
\epsfxsize=280pt{\epsffile{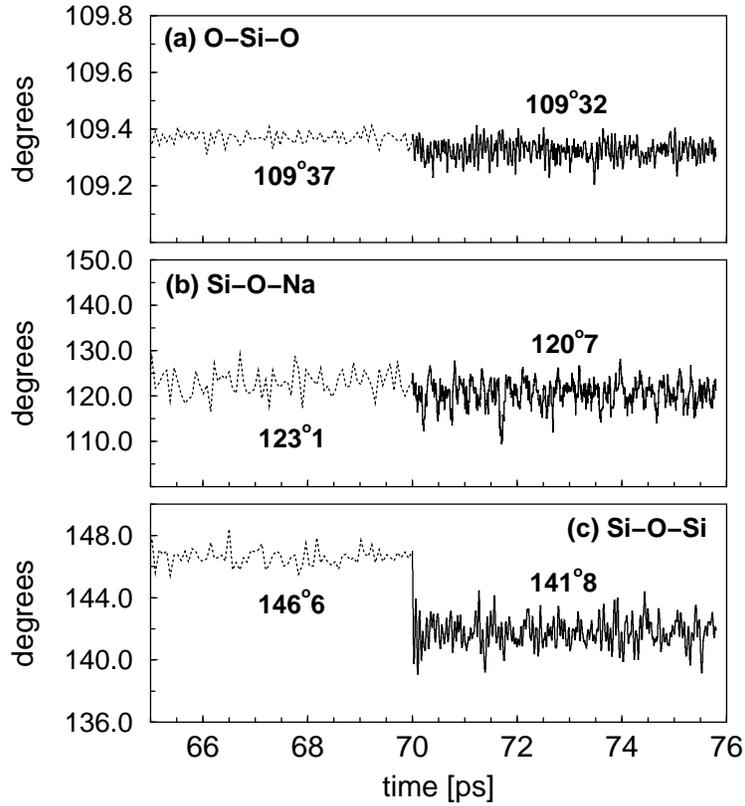}}
\caption{\label{evol-angles}  Time evolution of the mean O-Si-O (a), Si-O-Na (b) and 
Si-O-Si (c) bond angles during the last 5 ps of the classical 
MD simulation (dotted lines) and the full Car-Parrinello MD simulation (solid lines).}
\label{fig:evol-angles}
\end{figure}

First, as in the silica glass study \cite{sio2_cp_bks}, 
an important variation  occurs  in the mean value of 
the inter-tetrahedral angle Si-O-Si, which decreases from 146.1${}^{\mathrm o}$ 
to 141.8${}^{\mathrm o}$  -- see Fig. \ref{evol-angles}(c) 
which  shows  the  time dependence of the Si-O-Si mean bond-angle during the 
last 5 ps of the classical trajectory and the full CP trajectory.  Even if the behaviour 
was similar for both samples, we have chosen to present in 
Fig. \ref{evol-angles} only the data corresponding to  one sample. 
Similarly, in the remaining part of this subsection,  all the quantities discussed 
as a function of time have been plotted for one sample.
All the other characteristics have been averaged over both samples.

In Fig. \ref{evol-angles}(a) and (b), the time dependencies of the O-Si-O and  
Si-O-Na mean bond-angles during the same  time spans   are depicted. 
We can remark that both descriptions give almost identical mean values for the 
tetrahedral angle  O-Si-O, very close to the ideal 
tetrahedral angle 109.47${}^{\mathrm o}$. Concerning the Si-O-Na angle, we have 
computed it taking into account that each NBO has  two Na atoms in its nearest 
neighbourhood as mentioned above. Consequently the depicted values have been taken 
as the average value of the two corresponding Si-NBO-Na linkages.
By looking  at the time evolution of the mean value of this  angle,  we note firstly 
that it presents a slight shift to a lower value during the CP trajectory. But, since its   
fluctuations during the CP simulation are slightly larger compared to the BKS ones, 
we cannot say that the observed shift is a real one without knowing what the origin 
of these increasing fluctuations is. At  first glance, one  may  think  that 
these increasing fluctuations would be due to  the temperature shift.
 Hence, in order to check this hypothesis,  we have classically heated 
up one of the BKS samples to the CP temperature  (i.e. approximately  80 K higher) and 
let it relax during 70 ps at this new temperature. Neither the mean value of the  
Si-O-Na angle nor the amplitude of the fluctuations showed any real variation for 
the heated BKS sample.  In view of this result, we cannot interpret the increasing of 
the Si-O-Na angle fluctuations between the classical and the {\it ab initio} description as being 
a  temperature effect. They  can be much more probably due to the 
differences in the nearest-neighbour distances observed between the BKS and the CP 
descriptions (see below). On the other hand, these changes may indicate the fact that 
the ionicity of the Na-NBO bonds is better described by the CP approach. 
The Na-NBO bonds are therefore weaker than in the BKS description and this would induce 
larger fluctuations in  the Si-O-Na angle and equally in the Na-NBO mean distances, 
as we  will see in the next paragraphs.

\begin{figure}
\epsfxsize=280pt{\epsffile{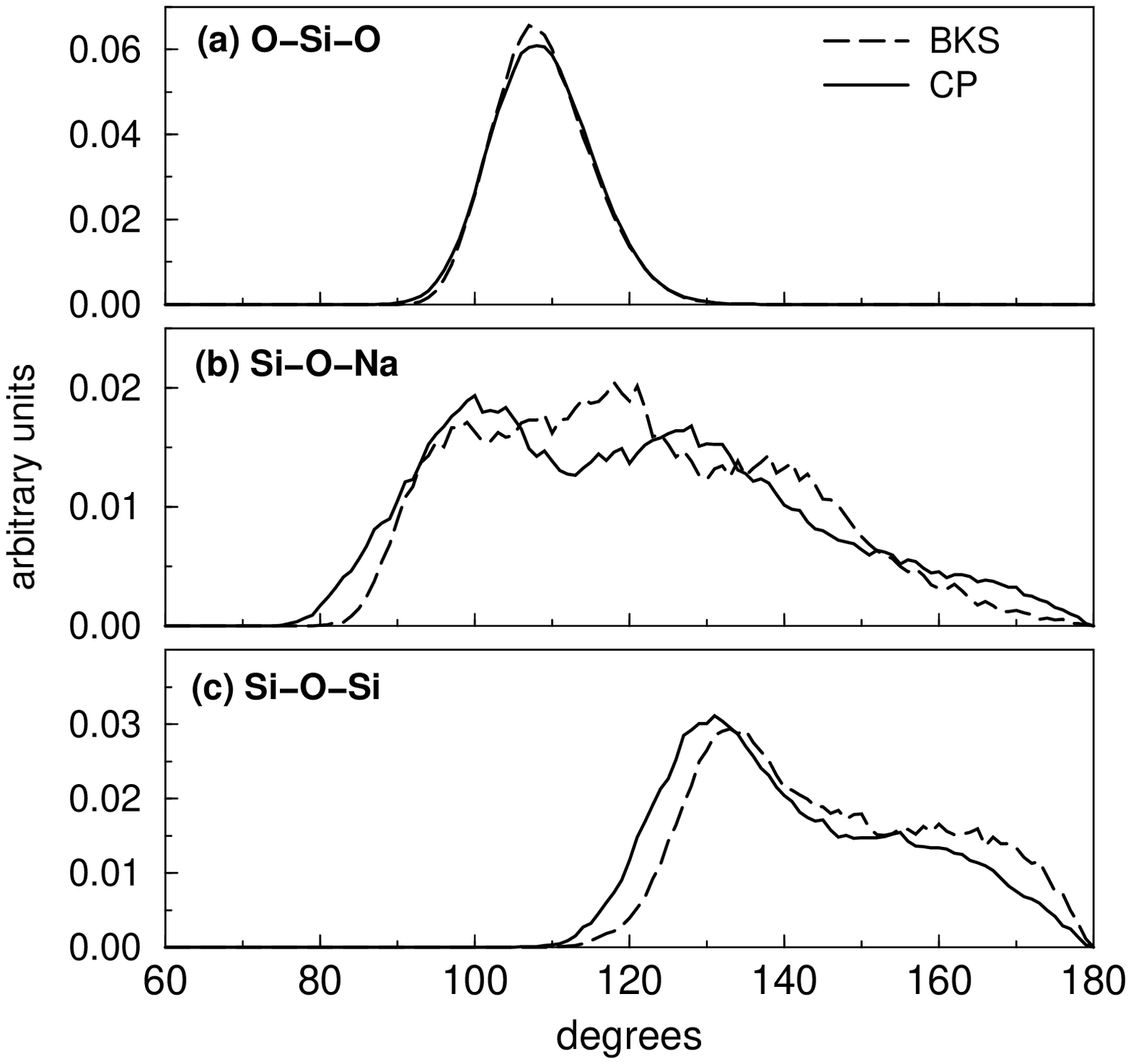}}
\caption{\label{distr-angles}  The O-Si-O (a), Si-O-Na (b) and Si-O-Si (c) time-averaged 
angle distributions from the Car-Parrinello MD simulation (solid line) and 
from the classical MD simulation (long dashed line), respectively.}
\label{fig:distr-angles}
\end{figure}

In Fig.  \ref{distr-angles} we have  reported the  time-averaged distributions of the 
tetrahedral O-Si-O and inter-tetrahedral  Si-O-Si angles  as well as that of the 
Si-O-Na bond angle, evaluated for both the BKS and CP trajectories. 
The O-Si-O angular distribution (Fig. \ref{distr-angles}(a))
shows in both cases a sharp peak at  $109{}^{\mathrm o}$ 
as for the pure silica and this demonstrates that the addition of the sodium oxide  
does not modify the tetrahedral environment of the Si atoms as was already pointed out 
from previous classical MD calculations \cite{huang_90,oviedo_98} and
 in a RMC fit of the  NS4 neutron diffraction data \cite{zotov_98_2}. 
For the Si-O-Na bond-angle (Fig. \ref{distr-angles}(b)), 
we note that both distributions have the same average 
values (i.e. $120^{\mathrm o}\pm 19$ in the BKS case and $120^{\mathrm o}\pm 22$ 
in the CP one), even if their shapes are not really identical.

From the comparison of the Si-O-Si distributions (Fig. \ref{distr-angles}(c)), 
it follows that the CP distribution presents a  shift towards lower values, 
already  noticed in the silica study \cite{sio2_cp_bks} 
and which is of course consistent with  the decrease of the mean angle mentioned above. 
Nevertheless both  distributions have an asymmetric shape as obtained 
in the recent RMC fits \cite{zotov_98_2} even if their 
average value   ($136{}^{\mathrm o} \pm 16$) is slightly lower than  those 
given by the present CP and BKS simulations (i.e.  $141{}^{\mathrm o} \pm 15$ in the CP case and 
$145{}^{\mathrm o} \pm 15$ in the BKS one). However, in the context of the controversy existing
in the literature on the Si-O-Si angle distribution for pure silica as well as for other silicates,
we note that the average values given here are  comparable with the range of results found 
up to now  by different groups and which lie between $133{}^{\mathrm o}$
and $160{}^{\mathrm o}$ \cite{greaves_81,huang_90,oviedo_98,zotov_98_2,mcdonald_67}.

\subsubsection{First neighbour distances, pair correlation functions and structure factor}

\begin{figure}
\epsfxsize=280pt{\epsffile{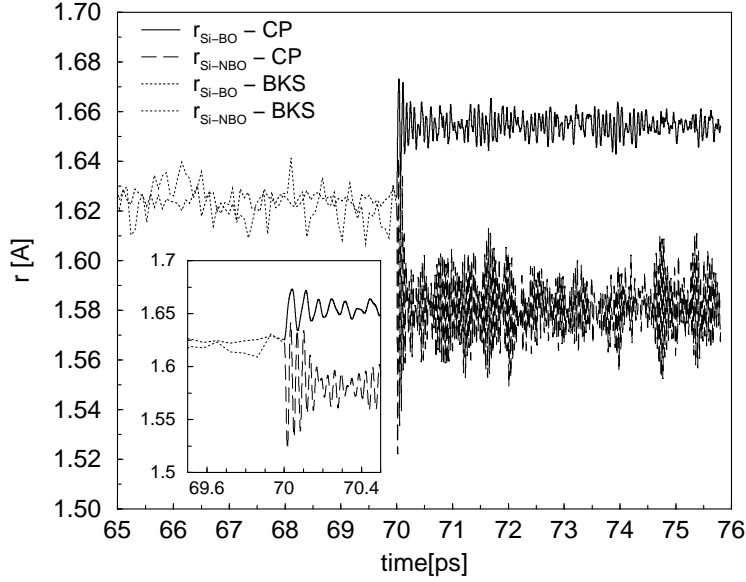}}
\caption{\label{si_bo-nbo}  Time evolution of the Si-BO and  Si-NBO 
  mean distances during the last 5 ps of the classical MD simulation and 
the full Car-Parrinello MD simulation.}
\label{fig:si_bo-nbo}
\end{figure}

The  second significant effect mentioned at the beginning of this section and which shows up  
as soon as  the CP simulation starts, concerns the evolution of the mean distances between 
the Si atoms and their oxygen nearest neighbours. Hence, once the CP dynamics starts, 
we have noticed, by comparison with the BKS results, an elongation of the mean  distances 
between the Si atoms and their BO  nearest neighbours and a shortening of the mean  
distances between the Si atoms and their NBO  nearest neighbours (see Fig. \ref{si_bo-nbo}). 
As can be seen in Fig. \ref{si_bo-nbo}, these mean distances for the BKS simulated glass 
have identical values ($\approx 1.62$ \AA ) while there exists an important splitting  
in the CP case which takes place after a very short relaxation (see the inset of the same figure) 
and remains remarkably stable during the whole CP trajectory (the CP Si-BO and Si-NBO mean 
distances are equal to  1.65 \AA \ and  1.58 \AA \ respectively).

\begin{figure}
\epsfxsize=280pt{\epsffile{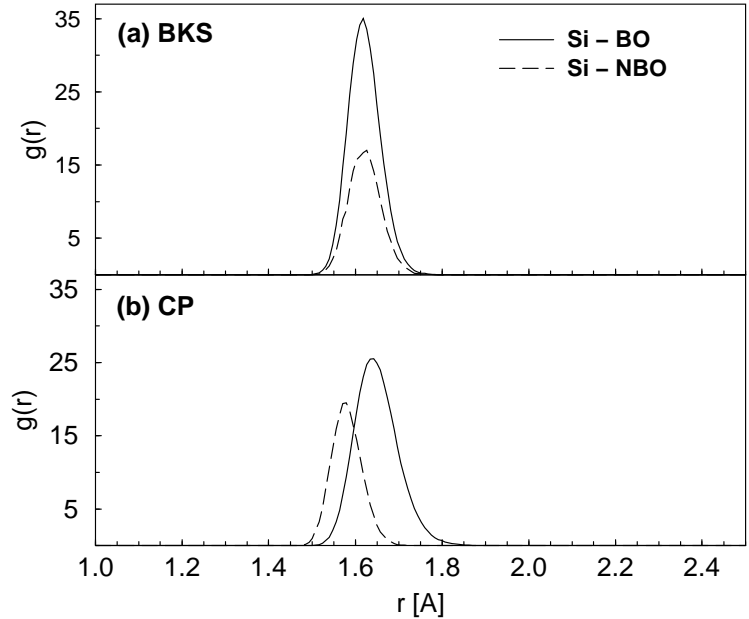}}
\caption{\label{gdr-si-bo-nbo}  First peaks of the Si-BO (solid line) 
and Si-NBO (dashed line) pair correlation functions of the NS4 glass 
computed from the classical MD  (a) and  the Car-Parrinello MD (b) simulations.}
\label{fig:gdr-si-bo-nbo}
\end{figure}

 The shortening of the Si-NBO distance and the lengthening of the Si-BO distance are also 
reflected by the Si-BO and Si-NBO pair correlation functions plotted in Fig. \ref{gdr-si-bo-nbo}. 
These results are in very good agreement with  MD \cite{huang_90} and RMC  \cite{zotov_98_2} 
studies  reported for the NS4 glass  as well as with the experimental and MD data reported 
for other crystalline and amorphous sodo-silicates \cite{melman_91,mcdonald_67}. 
One has to remark that, in the quoted MD simulations \cite{huang_90,melman_91} the authors 
have used three body potentials while the potential used in the present BKS simulation 
is a pair potential one. This may be the reason why the quoted MD calculations exhibited
different values for the Si-BO and Si-NBO mean distances.

\begin{figure}
\epsfxsize=280pt{\epsffile{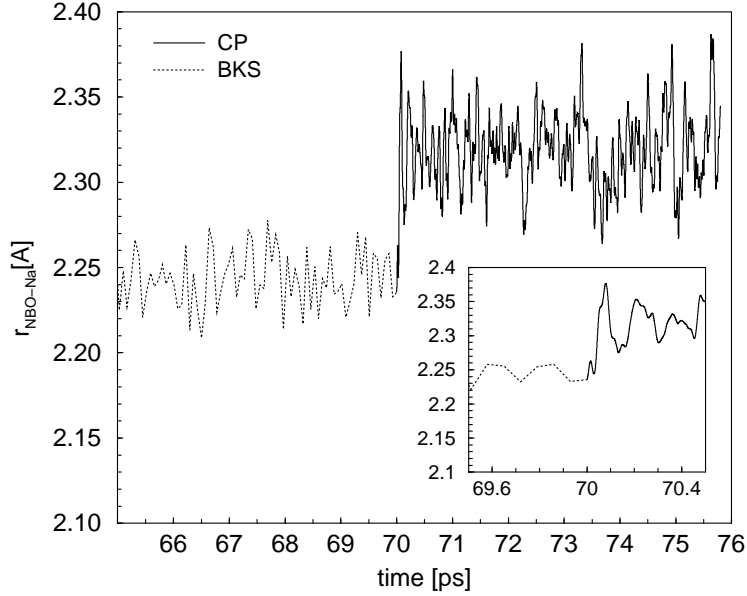}}
\caption{\label{na_o}  Time evolution of the NBO-Na mean distances during 
the last 5 ps of the classical MD simulation (dashed line) and the full Car-Parrinello 
MD simulation (solid line).}
\label{fig:na_o}
\end{figure}

We have also investigated what happens with  the mean distance between a NBO atom 
and its Na nearest neighbour. As shown in Fig. \ref{na_o}, it increases once the
CP simulation begins. We recall that two Na atoms have been identified in the nearest 
neighbourhood of each NBO atom and the values given in Fig. \ref{na_o} are the averages 
of the two NBO-Na distances. The lengthening of this distance which occurs also very 
shortly after the plugging in of the CP description reflects, together with the effects 
described above, the ability of the CP description to change the local 
environment around the oxygen atoms even at low temperature.

\begin{figure}
\epsfxsize=280pt{\epsffile{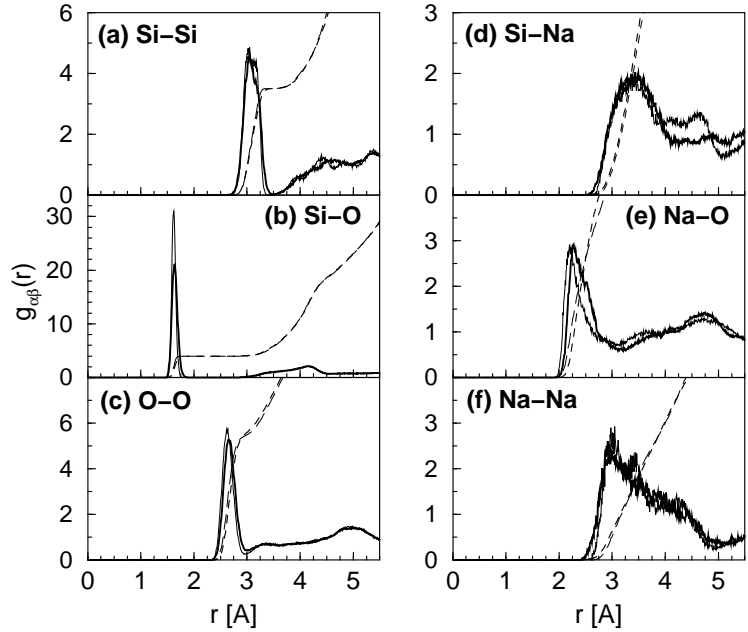}}
\caption{\label{gdr_fig}  Pair correlation functions of the NS4 glass obtained 
from the Car-Parrinello MD (bold solid lines) and  classical MD (thin solid lines)
 simulations. The bold dashed and the thin long-dashed lines are the corresponding integrated 
coordination numbers. }
\label{fig:gdr_fig}
\end{figure}

The pair correlation functions (PCF)   $g_{\alpha ,  \beta} (r)$ ($\alpha , \beta =$ Si, O, Na) 
and the corresponding integrated coordination numbers (CN) are shown in Fig. \ref{gdr_fig} for 
both BKS and CP simulations. We note that the addition of the  sodium oxide does not change 
the positions of the peaks  involving the network forming atoms (Si-Si, Si-O and O-O, 
Fig. \ref{gdr_fig}(a), (b) and (c)) which remain
almost the same as those found for pure silica and the two descriptions give overall similar 
results. Hence the first peaks in the Si-O correlation function exhibit maxima  at values of 
$r$ close to $\approx$ 1.62 \AA ,  the corresponding experimental value calculated from EXAFS 
spectra \cite{henderson_95} being 1.617  \AA . The Si-O CN being equal to 4, there is no doubt 
that the SiO${}_4$ tetrahedron remains the basic unit in spite of the presence of the alkali oxide. 

When we compare the PCF involving the modifier cation, we note some differences between the results 
of the BKS and the CP simulations. Firstly, from the analysis of the Na-O PCF (Fig. \ref{gdr_fig}(e)), 
it follows that the BKS Na-O first peak is located at  $\approx$ 2.2 \AA \ while the CP one at 
$\approx$ 2.3 \AA.  This shift of the average first neighbour Na-O distance towards higher 
values of $r$ in the CP description is of course consistent with the increase observed in 
the time evolution of the NBO-Na mean distance (see Fig. \ref{na_o}). The CP average first 
neighbour Na-O distance  shows a very good agreement with previous XAFS measurements 
\cite{houde-walter_93,greaves_95} which gave an average value of 2.32 \AA . 

Concerning the Si-Na PCF (Fig. \ref{gdr_fig}(d)), we can notice that its shape is not similar  
for the classical and {\it ab initio} descriptions. After  an almost identical first   
peak located at $\approx$ 3.4 \AA, the BKS function presents a kind of plateau for $r$ varying from 
$\approx$ 4 to 4.8 \AA.  During the CP simulation, one can see that this plateau disappears.
Since this plateau is more pronounced in one of the two BKS samples,  it could simply be 
an artefact of the unrealistic quench rates that one has to impose usually in MD simulations. 
On the other hand, this plateau has been observed also in a classical MD simulation 
using the same BKS potential for one NS4 sample containing 648 atoms \cite{sunyer_jund} which shows
that it is not due to finite size effects.  The first peak position obtained here is 
in agreement with the RMC \cite{zotov_98_2} fits and previous classical MD results 
\cite{huang_90,oviedo_98}, but it is lower than that  estimated from sodium K-edge 
XAFS experiments  (i.e. 3.8 \AA) \cite{houde-walter_93,greaves_95}. 

From the same XAFS experiments \cite{houde-walter_93,greaves_95}, the Na-Na
average distance have been refined at 3.2 \AA. In the present work, the first peak of 
the Na-Na PCF has a maximum at $\approx$ 3 \AA \ for both BKS and CP case (see Fig. \ref{gdr_fig}(f)).
In the BKS case, the  Na-Na  PCF shows a splitting which seems to be smoothed in the CP case.
 However, it is difficult to extract really useful information concerning the Na-Na correlation
 given the poor statistics which is due to the small number of Na atoms in the system.

\begin{figure}
\epsfxsize=280pt{\epsffile{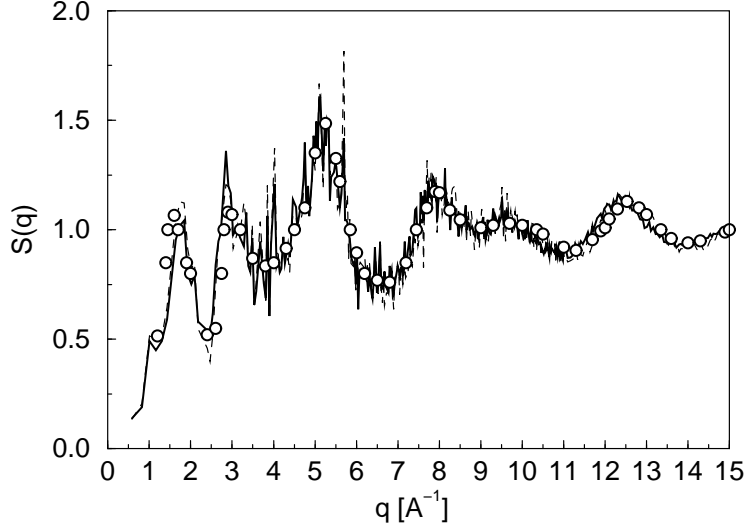}}
\caption{\label{sq_fig}  Comparison between the experimental 
neutron diffraction structure factor (the circles) that we have 
extracted from reference \protect\cite{zotov_98_2} and the computed static 
structure factors from Car-Parrinello MD (solid line) 
and classical MD (dotted line) simulations.
For the calculations of the structure factors, 
the scattering lengths $b_{Si}=$ 4.149 \AA\ , $b_O=$ 5.803 \AA\ 
and $b_{Na}=$ 3.63 \AA\ were used.}
\label{fig:sq_fig}
\end{figure}

Figure \ref{sq_fig} shows the total static structure factor computed from the CP and BKS 
simulations compared with the experimental points that we have extracted from a  neutron 
diffraction experiment \cite{zotov_98_2}. The differences between the curves corresponding 
to the two simulations are indistinguishable and they agree reasonably with experimental data. 
Nevertheless it is interesting to notice that the microscopic discrepancies between the two 
descriptions, reported above, are totally smoothed out in the total structure factor. 
Thus, the good agreement shown by the $S(q)$ is not sufficient to draw conclusions on 
the validity of a model and a more detailed analysis  of the structure is required.

\subsection{Electronic properties}
\label{sec:electronic}

In this subsection, we  analyze  the electronic properties of the  two NS4 {\it ab initio}
 samples in terms of the electronic density of states and the Hirshfeld charge variations. 
These properties were computed for the final configurations of the CP runs 
 and compared to the ones obtained for the initial configurations (corresponding 
to the final configurations of the classical BKS simulations).
We compare the corresponding NS4 results with those obtained
for the {\it ab initio} silica glass sample  described in Ref. \cite{sio2_cp_bks}. 

\subsubsection{Electronic density of states}

\begin{figure}
\epsfxsize=280pt{\epsffile{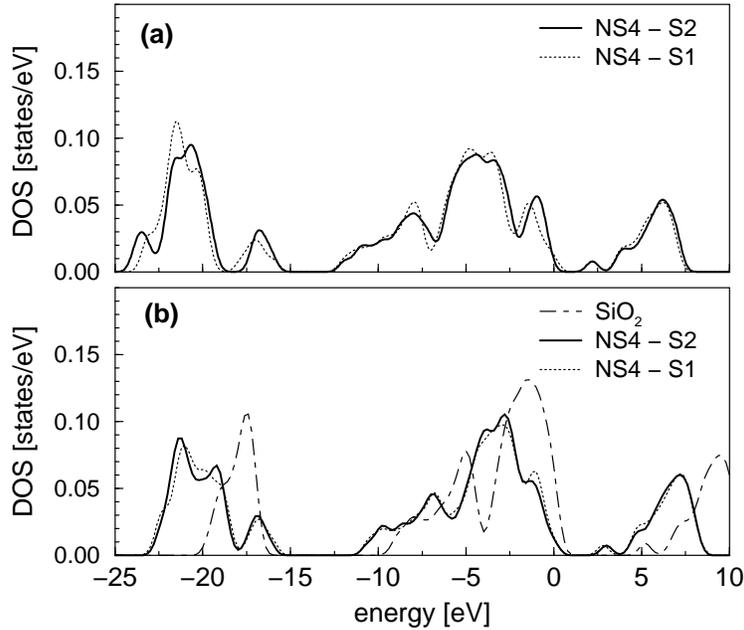}}
\caption{\label{DOS} Upper graph:  Kohn-Sham electronic density of states of the initial CP configuration of the two
NS4 samples at $\approx$ 300 K, corresponding to the last configuration of the BKS simulations,  
denoted S1 and S2 (dotted and solid lines). 
Lower graph: Kohn-Sham electronic density of states of the last CP configuration of the two NS4 samples 
at $\approx$ 400 K, denoted  S1 and S2 (dotted and solid lines) compared to the SiO$_2$ density of 
states of Ref. \protect\cite{sio2_cp_bks} at $\approx$ 335 K (dotted-dashed line). 
The Fermi levels have been shifted to zero and the curves have been smoothed with a 
Gaussian broadening of same width.}
\label{fig:DOS}
\end{figure}

The Kohn-Sham (KS) eigenvalues are computed for both NS4 samples 
and the resulting electronic density of states (DOS) is depicted in Fig. \protect\ref{DOS}(a) for the initial
configurations of the CP runs, and in Fig. \protect\ref{DOS}(b) for the final ones. 
In Fig. \protect\ref{DOS}(b), the KS densities of states for NS4 are compared 
to the density of states of the silica glass sample studied in Ref. \cite{sio2_cp_bks}.
However, since the present NS4 study has been carried out using the LDA-B-LYP functional for the
exchange and correlation energy term, we have also computed the KS eigenvalues of the silica glass
using the same approximation (these eigenvalues were originally computed 
within the LDA description in Ref. \cite{sio2_cp_bks}).
 We have found only very slight differences between the SiO$_2$ DOS 
using  LDA and LDA+B-LYP. At $\approx$ 335 K, the band gap is found to be equal to 5.01 eV with the 
LDA+B-LYP description, which is much lower than the experimental value for amorphous silica ($\approx$ 9 eV).
This underestimation of the band gap 
is a well-known failure of DFT in the local density approximation \cite{dft_gap}.

The band gaps obtained for the two NS4 samples at $\approx$ 400 K are equal to 2.77 eV and to 2.86 eV,
respectively (Fig. \protect\ref{DOS}(b)). For comparison, the band gaps obtained for these two samples
in their initial configurations were equal to 2.16 eV and 2.17 eV, respectively 
(Fig. \protect\ref{DOS}(a)).  Hence, the refinement of the NS4 glass structure by 
the Car-Parrinello dynamics leads to an increase of the band gap.
By comparing with the value obtained for the silica glass, 
we note that the gap is significantly reduced by the introduction of Na atoms, 
as  had already been deduced from UV absorption experiments \cite{sigel,ab_initio1} 
and  suggested in  LCAO calculations \cite{edos}.

The DOS for the two NS4 samples are very similar and show common features with the
DOS of SiO$_2$ (Fig. \protect\ref{DOS}(b)). They mainly consist in oxygen $2p$ non bonding
orbitals (lone pair) for the highest occupied states, 
in bonding states between silicon $sp^{3}$ hybrids and oxygen $2p$ orbitals 
(from -10 eV to -4 eV for SiO$_2$ and from -11 eV to -5.4 eV for NS4) and in oxygen  
$2s$ orbitals for the lower energy band. The empty band is mainly derived from
weak anti-bonding conduction states. 
Note that in our NS4 samples, we take explicitly into account semi-core electrons 
 for the Na atoms. These electrons give rise to a low energy band at $\approx$ -48 eV which 
has not been considered in Fig. \protect\ref{DOS}.

Murray {\it et al.} \cite{edos} computed the DOS for sodium silicate
model glasses with various compositions, using a LCAO computational method.
Since they used atomic orbitals in their model, they could assign different 
contributions in the bands to states belonging to specific atoms. 
For instance, the NBO atoms gave rise to subpeaks or shoulders  to the right
of the main O $2s$ and O $2p$ bands. 
In the Kohn-Sham description, the orbitals are not atomic orbitals and consequently 
such an analysis of the bands is not possible. However, from the comparison of our DOS and the
DOS reported in Ref. \cite{edos}, we note that they are surprisingly similar and 
therefore, we can attribute the peak 
just below the Fermi level and the peak at $\approx$ -16.5 eV to the NBO atoms. 
In Ref. \cite{bruckner}, the reported value for the 
chemical shift between the BO and the NBO 1s-photoelectron bands 
measured by XPS experiments was about 2.45 eV. 
We find values of $\approx$ 2.5 eV for the O $2s$ bands and of $\approx$ 2.0 eV for
the O $2p$ bands. 

By comparison, the bands for the initial configurations - corresponding to the structure
given by the BKS potential (Fig. \protect\ref{DOS}(a)) - show only very slight differences 
with the bands
computed at the end of the CP runs. The O $2s$ bands are slightly modified and 
the peak just below the Fermi level attributed 
to the NBO atoms seems to be shifted towards lower values after the CP structure refinement.
This is consistent with the Si-NBO bond length shortening that occurs after the CP is switched on.

\subsubsection{Atomic charges}

It is well-known that the charge of an atom can never been defined uniquely and
it is not subject to experimental measurement. Nevertheless, it is possible to
use approaches such as Mulliken population analysis \cite{Mulliken} or  Hirshfeld 
determination of the density deformation \cite{hirshfeld} in order 
to discuss bonding or to compare the electron distributions in different systems 
using the same description. 
Here we have chosen to use the Hirshfeld description to determine variations of the
atomic charges between the NS4 samples and the silica samples. 
In this scheme, the total electronic charge of a bonded atom is given by:
\[ {\cal Q}_i = - \int \delta \rho_i(r) dv \]
where $\delta \rho (r)$ is the atomic deformation density, defined as the
difference between what Hirshfeld calls the "charge density of the bonded atom" and the 
"atomic density", i.e. the density difference between  the free atom and the
bonded atom (see Ref. \cite{hirshfeld} for a more complete definition). 
Adding the charge of the nucleus $Z_i$ gives the net atomic charge: 
\[q_i = {\cal Q}_i + Z_i . \]
This description,
based on the integration of the atomic deformation densities, has been used to compute 
atomic charges \cite{hirshfeld} that appear reasonable in magnitude and show variations
consistent with the accepted ideas about electronegativity differences between 
atoms and groups. 
However, the absolute values obtained using this approach tend
to be smaller than values obtained by other methods, such as Mulliken analysis,
and only variations of these charges have been examined here. 
The calculations have been performed at 0 K with fully optimized
structures in the LDA+B-LYP description and the 
results are presented in Table \ref{table_charges}.

By comparing the Hirshfeld net charges  obtained in the two NS4 samples, we first
notice that the values are extremely similar, the differences being within the error bars.
We have also compared these charges with the ones computed on the initial configuration of 
the two CP runs and we have observed no significant differences, given the large
error bars.

In the NS4 glasses, the charges on the NBO atoms are, on average, 
more negative than the ones on the BO atoms, which is a generally admitted result. 
Moreover we note that the differences between the charges on the silicon atoms depend
on the character of the tetrahedra (Q$_n$). The positive charge on the silicon 
atoms decreases as  $n$ decreases, i.e. the proximity of the NBO and of the Na 
atoms has a direct effect on the silicon atom and thus on the tetrahedra. 
This effect is known as a "non-localized 
effect" of the Na atom on the silica network \cite{ab_initio1,bruckner,ab_initio2} and 
 has already been shown in {\it ab initio} molecular orbitals 
calculations on sodium silicate clusters \cite{ab_initio1,ab_initio2}.
 But even if the electronic properties of the clusters were in agreement 
with experimental results from  UV absorption and XPS measurements
 \cite{sigel,bruckner}, some of the geometrical features were not correctly  
reproduced (in particular NBO-Na bond lengths). 
In our simulation, this "non-localized effect" gives rise to
the increase of the Si-BO bond length and to the decrease of the Si-NBO one 
(as presented in subsection \ref{sec:struct}), while the Na-NBO bond 
length is in agreement with experimental values.
We also observe that the increase of the Si-BO bond length is more
pronounced in Q$_3$ and Q$_2$ tetrahedra than in Q$_4$ tetrahedra,
in parallel with the increase of the Si net charges.

On the other hand,  comparison of the average charges found on the BO atoms 
between the NS4 samples and the SiO$_2$ sample shows that  the BO atoms 
in SiO$_2$ have a more negative charge than the BO atoms in NS4. 
As a direct consequence, the Si atoms in SiO$_2$ are less  charged
than the corresponding Si atoms in the Q$_4$ conformation in the NS4 samples. 
This effect is also very probably a consequence  of the highly non-localized 
effect of the Na atoms on the silica network.


\section{Conclusion}
\label{sec:concl}

We have presented a structural and electronic analysis
 of a sodium tetrasilicate (NS4) glass by using a combination of classical and 
{\it ab initio} molecular dynamics simulations. The electronic structure has been 
computed in the framework of the Kohn-Sham density 
functional theory within the generalized gradient approximation using a B-LYP functional. 
To our knowledge, this is  the first study of a  binary silicate glass carried out in 
the frame of first-principles molecular dynamics calculations. 

The scheme  employed for the preparation of the {\it ab initio} NS4 samples   
economizes a considerable amount of computation time as was already 
shown in recent work on the structural and electronic properties of a 
silica glass \cite{sio2_cp_bks}. 
As in the silica study, the CP dynamics has presented a remarkable stability. 
In order to validate  this  approach, we have performed it twice and we have obtained  
the same behavior both times.

The structural features of the NS4 samples have been studied in terms of 
pair correlation functions, bond angle distributions, Q-species, bridging 
to non-bridging oxygen ratio and structure factor. We note that, once the CP dynamics 
was plugged in, some important structural changes occurred. Hence the shortening 
of the Si-NBO distance, the lengthening of the Si-BO and Na-NBO distances appearing 
immediately after the start of the CP simulation are in perfect agreement with 
the experimental data. These results validate our preparation method and show the 
ability of the CP description to refine the local environment of the atoms even 
at low temperature.

The electronic structure of the {\it ab initio} samples thus obtained has been 
analyzed in terms of the density of states and the atomic charge variations.
The results show that the introduction of sodium atoms into the silica network 
lowers the band gap and that it has a highly non-localized effect on the charges 
of the atoms in the network.

In conclusion, we have generated - at low computational cost - 
a fully {\it ab initio}  sodium tetrasilicate glass  which shows  
structural and electronic characteristics in very good agreement with experimental 
results.  This work is a first step towards accurate studies of detailed 
electronic and vibrational properties of the sodium tetrasilicate glass.


\acknowledgments

We would like to thank   Walter Kob and Mark Tuckerman
for very interesting and stimulating discussions and 
J\"urg Hutter for his invaluable  help with the CPMD code. 
The simulations have been performed on the IBM/SP2 and IBM/SP3 
in CINES, Montpellier FRANCE.


\newpage

\newpage

%
\begin{table}
\noindent\begin{tabular}{|l||c|c||c|}
 & \multicolumn{2}{c||}{Na$_2$O-4SiO$_2$} & SiO$_2$ \\
\hline  
 & sample 1 & sample 2  & sample from Ref \cite{sio2_cp_bks}  \\
\hline \hline 
BO   & -0.089 $\pm$ 0.011 & -0.089  $\pm$ 0.009  &   -0.109 $\pm$ 0.007  \\
NBO  &  -0.245 $\pm$ 0.020  & -0.249  $\pm$ 0.015  &    -      \\
Na   &   0.076 $\pm$ 0.038  &   0.082  $\pm$ 0.034  &    -      \\
Si (all atoms) &  0.240 $\pm$ 0.035  &  0.240 $\pm$ 0.041  & 0.218 $\pm$ 0.010  \\
Si ($Q_4$) &   0.262 $\pm$ 0.017  &  0.270 $\pm$ 0.012 &   0.218 $\pm$ 0.010  \\
Si ($Q_3$) &  0.222  $\pm$ 0.012  & 0.212 $\pm$ 0.022  &   -      \\
Si ($Q_2$) &  0.125  &  0.153 $\pm$ 0.005 &   -      
\end{tabular}
\caption{\label{table_charges}{Average Hirshfeld atomic net charges computed from
the fully optimized structures of the two NS4 samples  and  of the SiO$_2$ sample 
of Ref. \protect\cite{sio2_cp_bks} at 0 K. }}
\end{table}

\end{document}